\documentclass[%
preprint,
 amsmath,amssymb,
 aps,
]{revtex4-2}

\usepackage{graphicx}
\usepackage{dcolumn}
\usepackage{bm}

\begin{document}

\preprint{Physical Review Fluids}

\title{Peristaltic pumping in sub-wavelength channels}

\author{Jessica K. Shang}
\email{j.k.shang@rochester.edu}
\author{J. Brennen Carr}%
\author{Caroline D. Cardinale}
\author{Delin Zeng}
\affiliation{%
 Department of Mechanical Engineering, University of Rochester, Rochester, NY
}%

\date{\today}

\begin{abstract}
We apply the lubrication approximation to solve for the flow generated by a peristaltic traveling wave in a finite, planar channel, and examine the effect of channel length. Cerebrospinal fluid (CSF) is hypothesized to be peristaltically transported by arterial pulsations through the perivascular spaces in the brain. Previous studies of peristaltic perivascular models have chosen model lengths ranging from sub-wavelength, which is more physiologically realistic, to full wavelength.  Here, we solve for peristaltic flow rates for arbitrary lengths, and find that sub-wavelength channels significantly modulate the mean value, phase, and amplitude of flow rate for sinusoidal and general peristaltic waveforms. The boundary conditions create an internal pressure gradient such that the instantaneous flow rate varies along the length of the channel, and the difference between the ends and the middle of the channel is more pronounced for very short channels. This longitudinal distribution in flow rate is not observed \emph{in vivo} in perivascular spaces at the surface of the brain, and hence  sub-wavelength peristaltic models whose boundary conditions are isolated from the larger perivascular network are limited in their ability to reproduce perivascular flows.
\end{abstract}

\maketitle


\section{\label{sec:introduction}Introduction}

Cerebrospinal fluid (CSF) flows through the annular perivascular spaces (PVSs) surrounding the surface arteries of the brain. One hypothesis is that the flow is peristaltically driven by traveling-wave pulsations of the arterial walls, a mechanism termed ``perivascular pumping'' \cite{Hadaczek:2006}. The hypothesis has been supported by \textit{in vivo} measurements, which show an unsteady movement of CSF that is correlated with the cardiac cycle and modulated by varying arterial pulsation \cite{Bedussi:2017, Mestre:2018}.  

The perivascular network is a complex, bifurcating system, and its geometry and flow are not yet fully characterized. Hence idealized models of the PVS have been employed in theoretical and computational studies of perivascular pumping: channels \cite{Schley:2006}, annuluses \cite{Bilston:2003, Wang:2011,Asgari:2016,Kedarasetti:2020, Carr:2021}, and more recently, a single bifurcation \cite{Daversin:2020} (see review in Ref.  \onlinecite{Thomas:2019}). These studies have reached differing conclusions  on the feasibility of perivascular pumping. One variable between studies is the domain length. Because of periodicity, domains that are a single (or integer) wavelength in length are effectively infinitely long \cite{Jaffrin:1971, Wang:2011, Schley:2006}. In perivascular pumping, the wave amplitude is in microns and the wavelength is about 0.1 m, so the range of length scales is computationally expensive at integer wavelengths; the wavelength can be shortened to non-physiological values \cite{Bilston:2003, Carr:2021}. However, each PVS segment is shorter than a wavelength, $10^{-4}$ to $10^{-3}$ m  \cite{Bedussi:2017,Mestre:2018}. Simulations showed that the mean flow rate generated by peristalsis alone is orders of magnitude smaller in sub-wavelength models compared to integer-wavelength models, and the net flow rate could be increased to a physiologically relevant value by adding a  pressure difference across the model \cite{Kedarasetti:2020, Daversin:2020}.

In this study, we extend the lubrication approximation of peristalsis employed by others \cite{Shapiro:1969, Schley:2006} to finite-length planar channels with no applied pressure difference. While perivascular spaces are annular and not planar \cite{Tithof:2019}, the azimuthal and radial flows in peristalsis are small relative to the longitudinal flow \cite{Carr:2021}, consistent with the lubrication approximation, so we expect that the effects of domain length in 2D will also apply to 3D geometries. Finite wavelengths were previously studied in Ref. \onlinecite{Li:1993} for cylindrical tubes greater than a wavelength.  Our analysis is the first, to our knowledge, to analytically explore the regime where the domain is sub-wavelength in length, which is pertinent to perivascular flow. Our model demonstrates a significant effect of domain length on mean and instantaneous flow rate, and highlights spatiotemporal features of the flow that have not been noted in perivascular pumping models.

\section{\label{sec:mdodel}Mathematical Model}
\begin{figure}
  \centerline{\includegraphics[width=7.8cm]{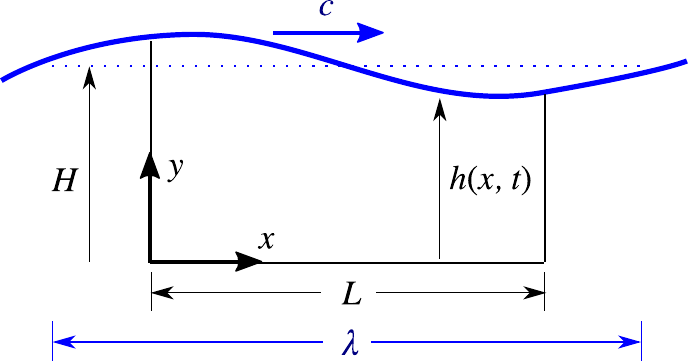}}
  \caption{Flow configuration.  Wave properties are shown in blue; channel properties are shown in black. As sketched, $L/\lambda < 1$.}
 \label{fig:setup}
\end{figure}

The channel is modeled as a two-dimensional, rectangular domain of height $H$ and finite length $L = n\lambda$, where $\lambda$ is the wavelength of the peristaltic wave and $n$ is a positive number of wavelengths, as shown in Fig. \ref{fig:setup}. The peristaltic wave propagates along the upper wall in the longitudinal $x$ direction, and its position is given by:
\begin{equation}\label{eq:heightfcn}
    h(x,t) = H(1+G(x-ct)),
\end{equation}
where $c$ is the wavespeed, $t$ is time and $G$ is an arbitrary waveform  that is continuous and periodic over the domain $x=[0, \lambda]$. The bottom wall is fixed. The channel is filled with an incompressible, Newtonian fluid of density $\rho$ and dynamic viscosity $\mu$, and is open, not porous \cite{MinRivas:2020}. The upper and lower boundaries are  impermeable. A pressure difference $\Delta p(t)$ can be applied across the channel; in this study, $\Delta p = 0$ to evaluate the effects of peristalsis alone. Non-dimensionalizing the Navier-Stokes equations yields an effective lubrication Reynolds number:
\begin{equation}
    Re_{\textrm{eff}} = \frac{\rho c H}{\mu} \cdot \frac{H}{\lambda}.
\end{equation}

\emph{In vivo} measurements at the surface of the brain of mice show that a typical PVS width is $H = 40\ \mu$m \citep{Mestre:2018}. The wavespeed $c$ is induced by arterial pulsations and is about 1 to 9ik10 m/s in mice \citep{Aslanidou:2016}; with a heartbeat frequency of 5 Hz, we calculate a  wavelength $\lambda = 0.1$ to 1 m. In humans, the PVS is a few hundred microns wide \citep{Bedussi:2017}, the arterial wavespeed is 1 to 10 m/s \citep {aPWV:2010}, and the heart rate is about 1 Hz, giving a wavelength $\lambda$ of 1 to 10 m. Cerebrospinal fluid (CSF) is similar in density and viscosity to water. Hence across a range of animal sizes, $H/\lambda \ll 1$ and $Re_{\textrm{eff}} \approx 10^{-3}$ to $10^{-2}$, which is sufficiently small, so we can apply the lubrication equations:
\begin{equation}\label{eq:lub}
    0 = -\frac{\partial p}{\partial x} + \mu \frac{\partial^2 u}{\partial y^2}; \ 0 = \frac{\partial p}{\partial y},
\end{equation}
subject to the boundary conditions:
\begin{eqnarray}
\label{eq:bc_u}u(y=0)=0, &&u(y=h) = 0, \\
\label{eq:bc_v}v(y=0) = 0, && v(y=h) = \frac{\partial h}{\partial t}, \\ 
\label{eq:bc_p}p(x = 0) &=& p(x=L),
\end{eqnarray}
where $u$ and $v$ are the velocities in the $x$ and $y$ direction, respectively, and $p$ is the pressure. We also note that the average Reynolds number calculated from experiments is also small, $Re = \frac{\rho \overline{u} H}{\mu} \approx 10^{-2}$, where $\overline{u}$ is the time-averaged velocity.

Our approach follows those in Refs. \onlinecite{Schley:2006},  \onlinecite{Shapiro:1969}, and \onlinecite{Li:1993}, and, like the latter, allows for non-integer $n$.  We solve Eq. \ref{eq:lub} for the velocity profile in the laboratory (fixed) frame:
\begin{equation}\label{eq:velprof}
u(x,y,t) = \frac{1}{2\mu}\frac{\partial p}{\partial x}y (y-h).
\end{equation}
We can  integrate continuity  $\frac{\partial u}{\partial x} + \frac{\partial v}{\partial y} = 0$
from 0 to $y$ and apply the boundary conditions for $v$ (Eq. \ref{eq:bc_v}):
\begin{equation}\label{eq:int_cty}
    \int_0^y \frac{\partial u}{\partial x} \ dy' + v(y)  = 0
\end{equation}
Substituting the velocity profile $u$, we obtain:
\begin{equation}\label{eq:pde}
 \frac{\partial}{\partial x} \left( \frac{h^3}{12\mu} \frac{\partial p}{\partial x}\right) = \frac{\partial h}{\partial t},
 \end{equation}
where any wall motion $h(x,t)$ instantly couples to a pressure adjustment inside the channel. Integrating with respect to $x$ and applying the pressure boundary conditions (Eq. \ref{eq:bc_p}) yields the pressure gradient:
\begin{equation}\label{eq:dpdx}
\frac{\partial p}{\partial x} = 12\mu h^{-3} \left(\int_0^{x} \frac{\partial h}{\partial t} \ d\zeta -\frac{\int_0^{L} \frac{\int_0^{\zeta} \partial h/\partial t \ d\xi}{h^3(\zeta,t)} \ d\zeta}{\int_0^{L} h^{-3}(\zeta,t) \ d\zeta} \right).
\end{equation}
With the pressure gradient, we can solve for the flow rate $q = \int^h_0 u \ dy$:
\begin{equation}
    q =  - \int_0^{x} \frac{\partial h}{\partial t} \ d\zeta + \frac{\int_0^{L} \frac{\int_0^{\zeta} \partial h/\partial t \ d\xi}{h^3(\zeta,t)} \ d\zeta}{\int_0^{L} h^{-3}(\zeta,t) \ d\zeta}.
\end{equation}

To simplify further, we can adopt a change of variables from $(x,t)$ to $z = x-ct$, where $z$ is in the traveling wave frame, and $h(x,t) = \hat{h}(z)$:
\begin{eqnarray}
    q &=&   \int_{-ct}^{x-ct} c\frac{d \hat{h}}{dz} \ dz + \frac{\int_{-ct}^{L-ct} \frac{\int_{-ct}^{x-ct} -c \ d\hat{h}/d\phi \ d\phi}{\hat{h}^3(z)} \ dz}{\int_{-ct}^{L-ct} \hat{h}^{-3}(z) \ dz} \\
    &=&   c(\hat{h}(z)-\hat{h}(-ct)) - c \frac{\int_{-ct}^{L-ct} (\hat{h}^{-2}(z) - \hat{h}(-ct)\hat{h}^{-3}(z)) \ dz}{\int_{-ct}^{L-ct} \hat{h}^{-3}(z) \ dz} \\
    &=& c\left(\hat{h}(z) - \frac{\int_{-ct}^{L-ct} \hat{h}^{-2}(z)  \ dz}{\int_{-ct}^{L-ct} \hat{h}^{-3}(z) \ dz}\right)
    \end{eqnarray}
and moving back into $(x,t)$, we obtain the flow rate in the laboratory frame:
\begin{equation}\label{eq:flux_general}
    q(x,t) = c\left(h(x,t)-\frac{\int_{0}^{L} {h}^{-2}(x,t) \ dx}{\int_{0}^{L} {h}^{-3}(x,t) \ dx}\right).
\end{equation}

Using the expression for height (Eq. \ref{eq:heightfcn}), the dimensionless flow rate expressed in terms of the displacement function $G$ is:
\begin{equation}\label{eq:flux_with_G}
    \frac{q}{cH} =  1+G(x,t)-\frac{\int_{0}^{L} {(1+G(x,t)})^{-2} \ dx}{\int_{0}^{L} {(1+G(x,t)})^{-3} \ dx}.
\end{equation}
The wave amplitude has been measured to be a few percent of the channel height \citep{Mestre:2018}, so $|G| \ll 1$. Expanded to second order in $G$,
\begin{equation}\label{eq:2nd_order_flux_with_G}
    \frac{q}{cH} \approx  1+G(x,t)-\frac{\int_{0}^{L} (1-2G + 3G^2) \ dx}{\int_{0}^{L} (1-3G+6G^2) \ dx}.
\end{equation}
In peristaltic pumping, we are interested in the mean flow rate, given by $\overline{q} = \frac{1}{\tau} \int^\tau_0 q(x,t) \ dt$, where $\tau = \lambda/c$. The dependence of the mean flow rate on the domain length $L$ is  entirely captured by the last term in Eqs. \ref{eq:flux_with_G} or \ref{eq:2nd_order_flux_with_G}.

\section{\label{sec:results}Results}

\subsection{Sinusoidal peristaltic wave}
To illustrate how the flow rate depends on the number of wavelengths $n$, we first consider a sinusoidal wave:
\begin{equation} 
G(x,t) = -\delta \sin \frac{2\pi}{\lambda}(x-ct)
\end{equation}
where $\delta$ is the dimensionless wave amplitude, normalized by $H$. The flow rate can be computed numerically from Eq. \ref{eq:flux_with_G} or be approximated by substituting $G$ in Eq. \ref{eq:2nd_order_flux_with_G} if $\delta$ is small. Retaining  oscillatory terms that are first order in $\delta$ after integration:
\begin{equation}\label{eq:q_approx}
    \frac{q(x,t)}{cH} \approx -\delta \sin \frac{2\pi}{\lambda}(x-ct) - \frac{C_0 + C_1 \sin\frac{2\pi}{\lambda}(c t-\frac{L}{2})}{1+D_1\sin\frac{2\pi}{\lambda}(c t-\frac{L}{2})},
\end{equation}
where 
\begin{equation}
    C_0 = -\frac{3}{2}\frac{\delta^2}{1+3\delta^2}, C_1 = \frac{\delta}{\pi n} \frac{\sin \pi n}{1+3\delta^2}, \ \textrm{and} \  D_1 = - \frac{3\delta}{\pi n} \frac{\sin \pi n}{1+3\delta^2}.
\end{equation}

\subsubsection{Average flow rate}
We obtain the time-averaged flow rate by integrating over one period $\tau = \lambda/c$:
\begin{equation}
\overline{q} = \frac{1}{\tau}\int^\tau_0 q(x,t) \ dt,
\end{equation}
which can be integrated exactly using the approximation for $q$ (Eq. \ref{eq:q_approx}). Retaining second-order terms in $n$ and $\delta$, the average flow rate is:
\begin{equation}\label{eq:mean_flux_approx}
    \frac{\overline{q}}{cH} \approx \frac{3}{2} \delta^2\left(1-\left(\frac{\sin \pi n}{\pi n}\right)^2\right).
\end{equation}
Thus the mean flow rate is independent of dimensional values such as the wavelength $\lambda$, and only depends on the number of wavelengths $n$ and amplitude $\delta$.

\begin{figure}
  \centerline{\includegraphics[width=12.6cm]{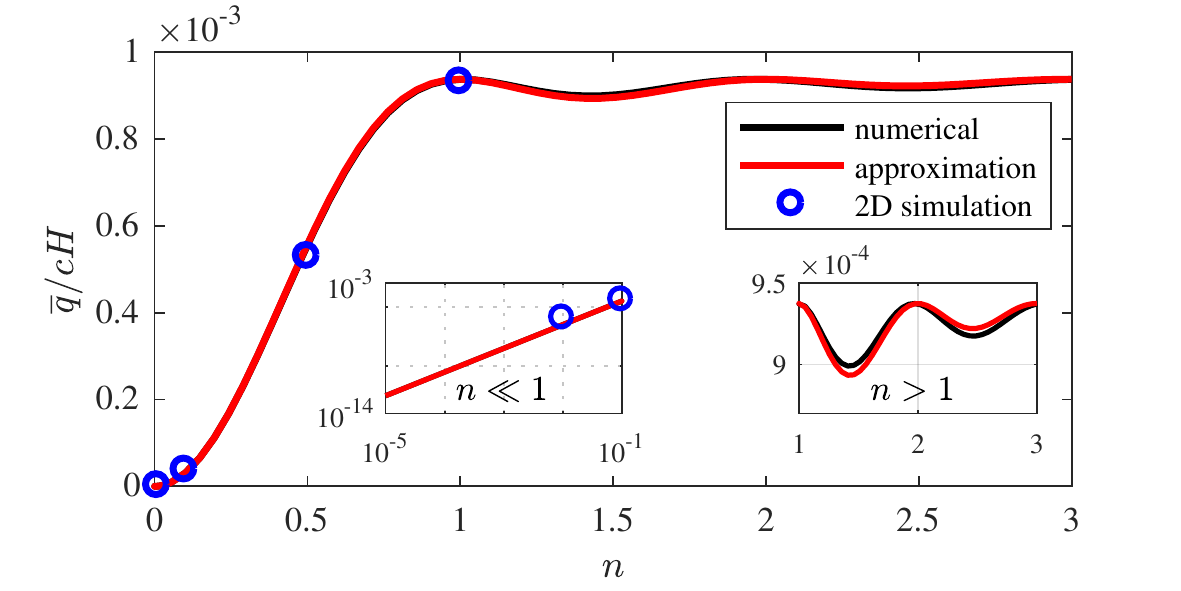}}
  \caption{Mean flow rate comparing the numerical solution of Eq. \ref{eq:flux_general} to the first-order approximation in Eq. \ref{eq:mean_flux_approx}, for $\delta = 0.025$. Inset details show mean flow rate at $n\ll 1$ and $n>1$. Values obtained with 2D finite-element simulations are also marked as validation; simulation details are included with the Supplementary Materials.}
 \label{fig:qbar_vs_n}
\end{figure}

This approximation for the  mean flow rate is compared with the numerical solution calculated with Eq.  \ref{eq:flux_with_G} for $\delta = 0.025$ in Fig. \ref{fig:qbar_vs_n}. As shown in the inset detail, the agreement is excellent  at the small values of $n$ characteristic of perivascular spaces. As expected, the mean flow rate varies periodically with $n$, and is maximized when  $n$ is an integer, i.e., infinitely long domains.  For non-integer $n>1$, the mean flow rate is slightly less than for integer $n$, and the difference diminishes with increasing $n$, which was also noted in Ref. \onlinecite{Li:1993} for finite tubes. For sub-wavelength domains, such as the $n<0.1$ lengths examined by previous investigators in different geometries \citep{Asgari:2016, Kedarasetti:2020, Daversin:2020}, the mean flow rate is several orders of magnitude smaller than for integer $n$. For  $n\ll 1$, a Taylor series expansion of Eq. \ref{eq:mean_flux_approx} in $n$ shows a quadratic dependence of $\overline{q}$ on $n$:
\begin{equation}\label{eq:mean_flux_approx_small_n}
    \frac{\overline{q}}{cH} \approx \frac{1}{2} \delta^2 (\pi n)^2,
\end{equation}
which is also seen in the inset in Figure \ref{fig:qbar_vs_n}.

\subsubsection{Dynamics}
\begin{figure}
  \centerline{\includegraphics[width=12.1cm]{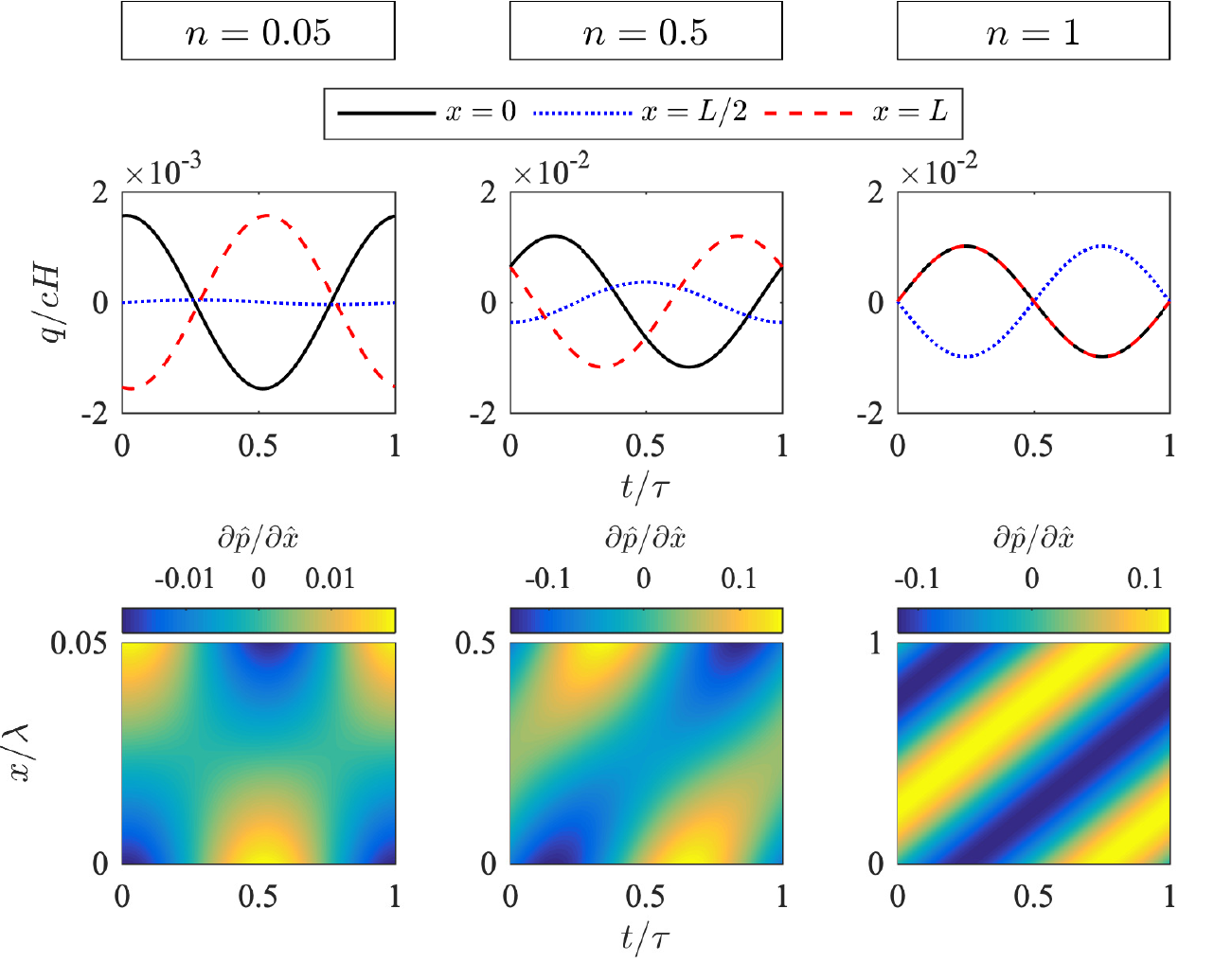}}
  \caption{Flow rate and pressure gradients for different channel lengths along the channel, for wave amplitude $\delta = 0.01$.}
 \label{fig:q_vs_t_var_n}
\end{figure}

For integer values of $n$, the flow rate $q(x,t)$ is in phase with $h(x,t)$: the last term in Eq. \ref{eq:flux_general} evaluates to a constant because the integrands are periodic. This also holds for our approximation in Eq. \ref{eq:q_approx}, since $C_1 = D_1 = 0$ for integer $n$. For non-integer $n$, the last term in Eq. \ref{eq:flux_general} varies with $x$ and $t$. In Eq. \ref{eq:q_approx}, $D_1$ is small for $\delta \ll 1$, which is generally the case in arterial pulsations. Neglecting this term, we can then approximate the oscillatory component of $q$ as:
\begin{equation}\label{eq:q_osc}
\frac{q_{osc}(x,t)}{cH} \approx  -\delta \sin \frac{2\pi}{\lambda}(x-ct) +\frac{\delta\sin\pi n}{(1+3\delta^2)\pi n} \sin \frac{2\pi}{\lambda}(\frac{L}{2}- c t).
\end{equation}
The second term is zero for integer $n$. We can further manipulate this into the form 
$q_{osc}(x,t)/cH = M \sin (\frac{2\pi}{\lambda}(x-ct)+\gamma)$ where $M$ is the amplitude and $\gamma$ is the phase shift, given by:
\begin{eqnarray}\label{eq:phase}
    \tan \gamma &=& \frac{\sin \frac{2\pi}{\lambda}(x-\frac{L}{2})}{\frac{(1+3\delta^2)\pi n}{\sin \pi n}-\cos \frac{2\pi}{\lambda}(x-\frac{L}{2})}, \\
    M &=& -\frac{\delta \sin \pi n}{\pi n (1+3\delta^2)}\frac{\sin \frac{2\pi}{\lambda}(x-\frac{L}{2})}{\sin \gamma}.
\end{eqnarray}
These expressions show that the phase shift and amplitude of the flow rate  vary along the channel when $n$ is not an integer. In Fig. \ref{fig:q_vs_t_var_n} (top row), we show $q(x,t)$ at different locations and $n\leq 1$. As expected, at $n=1$, $q$ has the same amplitude for all $x$, with no phase shift. In contrast, for smaller values of $n$,   $q$ varies with $x$ in amplitude and phase shift. The flow rate at the inlet and the outlet do not share the same phase as the driving wave. The dimensionless pressure gradient $\partial \hat p / \partial \hat x$ ($\hat p = pH^2/\mu\lambda c$, $\hat x = x/\lambda$), which can be computed from Eq. \ref{eq:dpdx}, is also shown in Fig. \ref{fig:q_vs_t_var_n} (bottom row). Recall that a uniform pressure is applied to either end of the channel, regardless of $n$. The wall motion generates an internal pressure gradient. For small values of $n$, the pressure gradient is about an order of magnitude smaller in amplitude compared to integer $n$, and is almost zero in the middle of the channel. Accordingly, the flow in the middle of the channel is very small, and increases in amplitude towards the ends of the channel.  As $n$ increases to 1, the amplitude of the pressure gradient and flow rate in the middle of the channel increases until their amplitudes match the those at inlet and outlet.

For small $n$, we illustrate some features in very short channels that will be helpful in our treatment of arbitrary waveforms. In this regime, $\sin \pi n \approx \pi n$, so the oscillatory flow rate (Eq. \ref{eq:q_osc}) can be rewritten as:
\begin{equation}
\frac{q_{osc}(x,t)}{cH} \approx  -\delta \sin \frac{2\pi}{\lambda}(x-ct) +\frac{\delta}{1+3\delta^2} \sin (\frac{2\pi}{\lambda}(x-ct)-\frac{2\pi}{\lambda}(x- \frac{L}{2})).
\end{equation}

At the inlet, $q_{osc}$ subtracts the wall displacement $G$ from another sine wave of almost-identical amplitude, differing by a small phase shift $\pi n$. Taking the difference between $\sin(z + \epsilon)$ and $\sin z$ effectively produces the derivative of the wave, i.e., the wall velocity, scaled by the factor $\epsilon$:
\begin{eqnarray}
\frac{q_{osc}(0,t)}{cH} &\approx&  -\delta \sin \left(\frac{2\pi}{\lambda}(-ct)\right) +\frac{\delta}{1+3\delta^2} \sin (\frac{2\pi}{\lambda}(-ct)+\pi n) \\
&\approx&  \pi\delta n \cos \frac{2\pi ct}{\lambda} \propto \frac{\partial G(0,t)}{\partial t}.
\end{eqnarray}
A similar result can be found for the outlet, which will have the opposite sign, as illustrated by Fig. \ref{fig:q_vs_t_var_n}. We can physically rationalize this wall-velocity-like behavior for small $n$. The local slope of the wall wave changes very slowly, $\delta \ll \lambda$, and the passing of the wall wave through short channels appears as a nearly uniform contraction and expansion of the channel height. For example, if the wave amplitude is 1\% of the channel height, then an estimate for the channel slope is $\delta/L \approx 0.01 H/(n\lambda)$; since $H/\lambda \ll 1$, then $\delta/L \ll 1 $. Since incompressibility is enforced, when the channel contracts, the fluid is squeezed out of either end, and when the channel expands, the fluid rushes back in. Hence the flow at the ends are in opposite directions and follow the wall velocity, and the middle of the channel experiences less flow, which was also observed in Ref. \onlinecite{Daversin:2020} in simulations of a finite bifurcation. 

\subsubsection{Mean flow versus oscillatory flow}

\begin{figure}
  \centerline{\includegraphics[width=11.5cm]{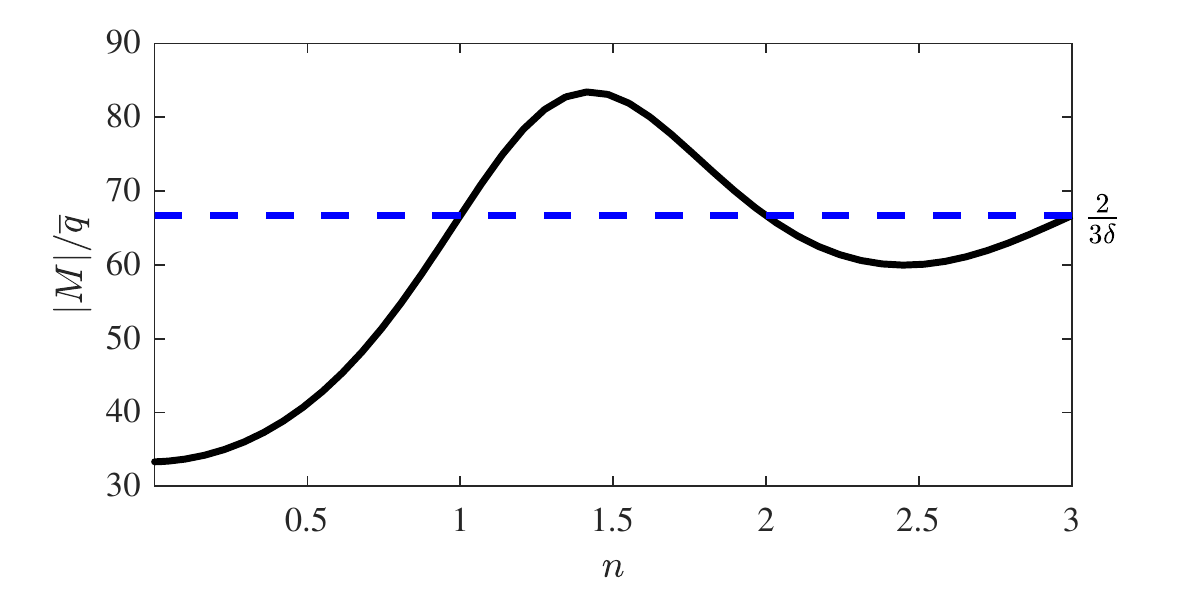}}
  \caption{Ratio of mean flow rate $\overline{q}$ to the amplitude of oscillatory flow  $|M|$ at the channel midpoint $x = L/2$, for $\delta = 0.01$, calculated from the numerical solution for Eq. \ref{eq:flux_general}.  For integer $n$, this ratio is $\frac{3\delta}{2}$, and for  $n \ll 1$, the ratio is $3\delta$, as marked in blue. }
 \label{fig:amp_ratio_vs_n}
\end{figure}

\emph{In vivo} experiments indicate that ratio of the time-averaged flow rate $\overline{q}$ to the oscillatory flow amplitude $|M|$ in the surface perivascular spaces in mice is about 2 \citep{Mestre:2018}. In the peristaltic model, Fig. \ref{fig:q_vs_t_var_n} shows that this ratio should be largest at the channel midpoint $x=L/2$ where the oscillatory flow is smallest. In Figure \ref{fig:amp_ratio_vs_n} we show the ratio as a function of channel length at this location. Evaluating Eqs.  \ref{eq:mean_flux_approx} and \ref{eq:q_osc}  at this location yields a ratio that is  $\frac{3\delta}{2}$ for integer $n$. Similarly, using Eqs. \ref{eq:mean_flux_approx_small_n} and \ref{eq:q_osc}, for $n\ll1$ representative of perivascular models, the ratio is approximately $3\delta$. Since this ratio is maximized at $x=L/2$, the ratio at other channel locations is at most O($\delta$) and always smaller than 1 for small wave amplitudes.

\subsection{General waveforms}
\begin{figure}
  \centerline{\includegraphics[width=12.1cm]{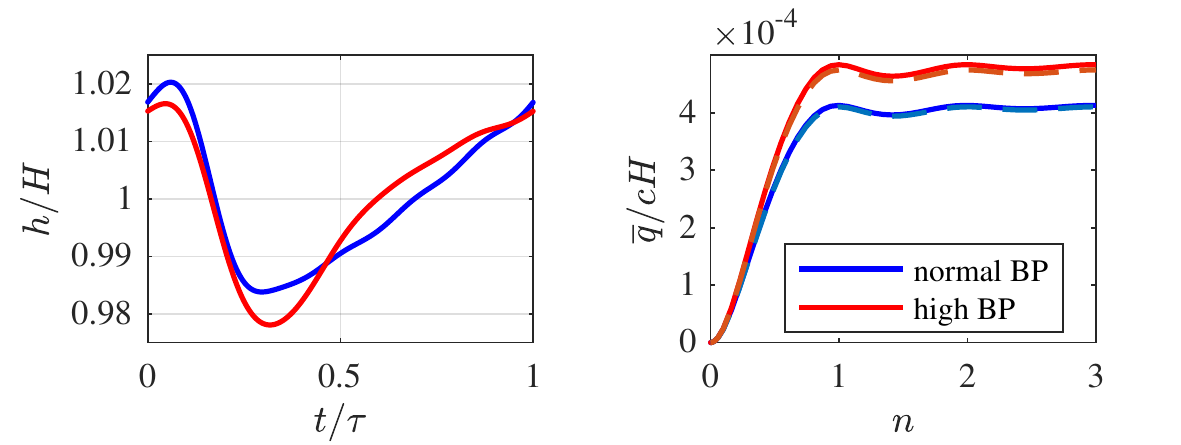}}
  \caption{Left: Fourier series representations of wall position, as measured by Ref. \onlinecite{Mestre:2018} for mice with normal and high blood pressure (BP). Right: Mean flow rate for the normal and high blood pressure waveforms, comparing the numerical solution (solid) against the approximate summation in Eq. \ref{eq:mean_flux_in_vivo} (dashed).}
 \label{fig:mestre_meanflux}
\end{figure}

A general waveform can be represented with a Fourier series:
\begin{equation}
    h(x,t)=H\left(1-\sum_{m=1}^{M} \delta_m \sin(\frac{2\pi m}{\lambda}(x-ct)+\phi_m)\right).
\end{equation}
The flow rate can be broken down into an oscillatory and mean component, $q(x,t) = q_{osc}+\overline{q}$. The flow rate is not linear in $h$, as illustrated by Eq. \ref{eq:flux_general}, so  we cannot generally sum the $M$  flow rates generated by each component in the series to obtain the total flow rate. If $|\delta_{m+1}|\ll|\delta_m|$, then the nonlinearity is reduced and the  flow rate can be approximated as:
\begin{equation}\label{eq:mean_flux_in_vivo}
    \overline{q}_{total} \approx \sum_{m=1}^{M} \overline{q}_m, \ \textrm{where} \ 
    \frac{\overline{q}_m}{cH} \approx \frac{3}{2} \delta_m^2\left(1-\left(\frac{\sin \pi mn}{\pi m n}\right)^2\right).
\end{equation}
In Fig. \ref{fig:mestre_meanflux}, we show Fourier series adapted to $\emph{in vivo}$ wall waves for normal and high blood pressure \citep{Mestre:2018}, with the mean flow rates calculated directly from Eq. \ref{eq:flux_general} and approximated with the sum in Eq. \ref{eq:mean_flux_in_vivo}. The agreement is good, since the $\delta_{m+1}\ll\delta_m$ condition is met. Since the amplitude is larger, more flow is predicted for the high blood pressure wave than for normal blood pressure. In contrast, Ref. \onlinecite{Mestre:2018} found that high blood pressure waves generated less net flow \emph{in vivo}.

\begin{figure}[t]
  \centerline{\includegraphics[width=12.6cm]{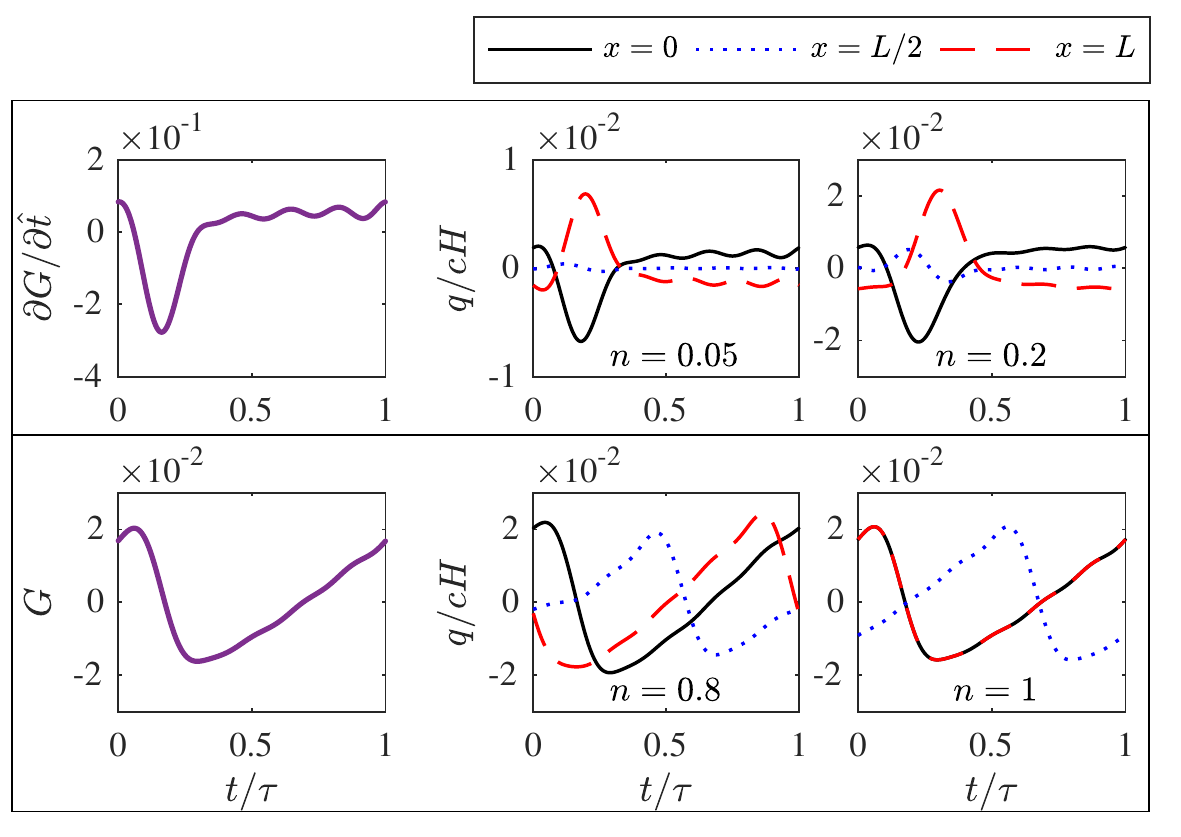}}
  \caption{Instantaneous flow rate for different $n$ generated by the normal blood pressure wave from Ref. \onlinecite{Mestre:2018}. For small $n$, the flow rate behaves like the wall velocity (top row), or like the wall displacement as $n\rightarrow1$ (bottom row).}
 \label{fig:mestre_q_vs_t_var_n}
\end{figure}

Fig. \ref{fig:mestre_q_vs_t_var_n} illustrates the effect of domain length on the instantaneous flow rate, using the normal blood pressure wave. Like the sinusoidal waves, for small values of $n$, we find that the inlet and outlet flow rates resemble the wall velocity, moving in opposite directions, as fluid squeezes out of the ends (top row).  In the middle, the amplitude decreases significantly as the pressure gradient becomes small. As the channel length increases, the velocity-like behavior fades, and the amplitude of the flow rate in the middle of the channel becomes larger. As $n$ increases to 1, the amplitude of the  second term in Eq. \ref{eq:q_osc} decreases, and hence $q(x,t)$ favors the wall displacement function $G(x,t)$ (Fig. \ref{fig:mestre_q_vs_t_var_n}, bottom). Refs. \onlinecite{Kedarasetti:2020} and \onlinecite{Daversin:2020} simulated the same wave function in 3D annuluses, and the reported flow rates at the end of their domains resembled those in Fig. \ref{fig:mestre_q_vs_t_var_n}, though they did not explicitly show the velocity profiles at other locations.

\section{Discussion and Conclusions}
In this study, we have examined the flow rate produced by a peristaltic wave in channels that are shorter than a wavelength, which is the regime where perivascular pumping is expected to take place.  We find that the mean flow rate depends only on the number of wavelengths $n$ and not on the physical wavelength $\lambda$, in agreement with the finite-tube solution of Ref. \onlinecite{Li:1993}. This implies that simulations with peristaltic waves that are shorter than physiological arterial waves are a reasonable model for perivascular pumping at whole wavelengths \citep{Bilston:2003, Carr:2021}, if the domain length and wave amplitude are scaled appropriately. The dependence only on $n$ and $\delta$ contradicts Ref. \onlinecite{Mestre:2018}, which attributed the decrease in mean flow rate in hypertension (high blood pressure) to a prolonged period of negative wall velocity. In Fig. \ref{fig:mestre_meanflux}, the high blood pressure waveform has a higher amplitude and resulting flow rate. Hence peristaltic motions are likely not solely responsible for the decreased flow rate seen in hypertension experiments.

The sensitivity of mean flow rate to $n<1$ agrees with previous studies that find that the mean flow rate at low $n$ is very small compared to that for domains of integer wavelengths \citep{Asgari:2016, Kedarasetti:2020, Daversin:2020}. We demonstrated that an internal pressure gradient generates flows that vary in phase shift and amplitude along the channel. At low values of $n$, the fluid is predominantly oscillated in and out of the channel, synchronous with the wall velocity, and the flow experiences much lower oscillations at the center of the channel. This wall velocity-like flow rate was observed in previous studies \citep{Asgari:2016, Kedarasetti:2020}, and Ref. \onlinecite{Daversin:2020} also noted that the flow rate amplitude increased towards the inlets and outlets of their bifurcation.  The \emph{in vivo} measurements made by Ref. \onlinecite{Mestre:2018} did not show significant changes in amplitude and phase between different positions along a segment of the PVS. Moreover, these measurements showed that the amplitude of oscillatory flow is comparable to the mean flow rate, whereas in our peristaltic model, the mean flow is significantly smaller than the oscillatory flow, dictated by the wave amplitude $\delta$.

To match physiological net flow rates in their simulations, Refs. \onlinecite{Kedarasetti:2020} and \onlinecite{Daversin:2020} prescribed a small static or pulsatile pressure difference across their models, since pressure gradients likely exist in the PVS. While we assume $\Delta p = 0$ in our study, an applied pressure difference $\Delta p(t)$ would add an unsteady term to Eq. \ref{eq:dpdx} which could also increase the  mean flow to oscillatory flow ratio. However, its contribution to the instantaneous flow rate would be independent of $x$, and so its effect is felt equally at all channel locations. The resulting flow rate still would be spatially inhomogeneous in phase and amplitude at small $n$, which is not observed \emph{in vivo}. A full-wavelength model is homogeneous, but as noted by Ref. \onlinecite{Kedarasetti:2020}, the flow amplitude is unrealistically large compared to the mean flow. A method to dampen the large oscillations in the PVS is to couple a Windkessel boundary condition to the PVS model to mimic the upstream and downstream vessel resistance and compliance \citep{Ladron:2020}.

In conclusion, the mean and spatiotemporal distribution of flow in a sub-wavelength peristaltic model is sensitive to its dimensionless length. Hence perivascular flow cannot be reproduced by a peristaltic model that considers a small subsection of the PVS with simple boundary conditions that do not capture the rest of the network and its surroundings. Additional mechanisms can be introduced to a short-domain peristaltic model that may be physiologically relevant for perivascular flow. Windkessel boundary conditions, as mentioned above, account for resistance and compliance outside of the model that can modulate amplitude and phase shift \citep{Ladron:2020}. Explicit inclusion of compliance or permeability in the outer wall of the PVS can introduce a local variation in pressure \citep{Kedarasetti:2020b, Romano:2020} that is not possible with the solid rigid wall used in the current peristaltic model. The development of complex boundary conditions in peristaltic models that can match observed perivascular flows is a rich area for future study, that can be improved in accuracy with expanded \emph{in vivo} measurements at multiple PVS locations of flow, wall displacement, and elusively, pressure.

\begin{acknowledgments}
We are grateful for constructive discussions with J. H. Thomas, D. H. Kelley, and J. Tithof.  This work was supported by the National Institute on Aging (RF1 AG057575-01), and the Office of Naval Research (N0001-18-1-2456).
\end{acknowledgments}

\bibliography{apssamp}

\end{document}